\documentclass[twocolumn,showpacs,amsmath,amssymb,superscriptaddress]{revtex4-1}
\usepackage{dcolumn}
\usepackage{color}
\usepackage{graphicx}
\usepackage{multirow}
\usepackage{amsmath}
\usepackage{bm}

\begin{document}

\title{Beyond mean-field boson-fermion model for odd-mass nuclei}

\author{K.~Nomura}
\author{T.~Nik\v si\'c}
\author{D.~Vretenar}
\affiliation{Physics Department, Faculty of Science, University of Zagreb, 10000 Zagreb, Croatia}

\date{\today}

\begin{abstract}
 
A novel method for calculating 
spectroscopic properties of medium-mass and heavy atomic nuclei with an odd 
number of nucleons is introduced, based on the framework of nuclear energy density functional theory 
and the particle-core coupling scheme. The deformation energy surface of the even-even core, as well as the 
spherical single-particle energies and occupation probabilities of the odd
particle(s), are obtained in a self-consistent mean-field calculation determined by the choice of the 
energy density functional and pairing interaction. This method uniquely determines the parameters 
of the Hamiltonian of the boson core, and only the strength of the particle-core coupling is 
specifically adjusted to selected data for a particular nucleus. The approach is illustrated in a 
systematic study of low-energy excitation spectra and transition rates of  
axially deformed odd-mass Eu isotopes. 
\end{abstract}

\pacs{21.10.Re,21.60.Fw,21.60.Jz}

\keywords{}

\maketitle


\section{Introduction}

As in many other quantum systems, the interplay between single-particle and
collective degrees of freedom plays a crucial role in the physics of atomic nuclei \cite{BM,RS,CasBook}. 
This is apparent especially in systems with odd number of protons
$Z$ or/and neutrons $N$. At low energies in nuclei with even $Z$ and $N$ 
nucleons are coupled pairwise and this is manifest in low-lying 
rotational and vibrational collective excitations \cite{BM}. 
Many nuclear models have successfully been applied in studies 
of the structure of even-even nuclei \cite{BM,RS,CasBook,caurier2005,IBM}. 
The situation is, however, more complicated in nuclei with odd 
$Z$ and/or $N$, because one has to consider unpaired fermions explicitly and treat
the single-particle and collective degrees of freedom on the same level \cite{BM,bohr1953}. 
Although most nuclear species have an odd $Z$ or/and $N$, microscopic studies of their structure have 
not been pursued as extensively as in the case of even-even systems, especially for medium-heavy and heavy nuclei.  

Nuclear density functional theory (DFT) provides a reliable global microscopic approach to 
many structure phenomena \cite{Ben03rev,Vre05,LNP.641,SR.07}. The basic implementation of the 
energy density functional (EDF) framework is in terms of
self-consistent mean-field (SCMF) methods, in which an EDF is constructed
as a functional of one-body nucleon density matrices that correspond to
a single product state. The static nuclear mean-field is characterized by the 
breaking of symmetries of the underlying Hamiltonian -- translational, rotational, particle 
number and, therefore, includes important static correlations, e.g. deformations and 
pairing. To calculate spectroscopic properties, such as excitation spectra and transition rates,
the mean-field approach has to be extended to include collective correlations that
arise from symmetry restoration and fluctuations around mean-field
minima. Collective correlations are taken into account through restoration of 
broken symmetries and configuration mixing of symmetry-breaking product states 
using, for instance, the Generator Coordinate Method (GCM) \cite{RS}. 

GCM configuration mixing of angular-momentum and particle-number projected states 
based on EDFs or effective interactions has become a standard tool for nuclear structure 
studies of even-even nuclei. However, considerable challenges are encountered when 
extending this method to odd-mass systems. 
In fact, it is only recently that such a consistent extension, where the generator 
coordinate space is built  from self-consistently blocked one-quasiparticle Hartree-Fock-Bogoliubov (HFB) 
states, has been reported in Ref.~\cite{bally2014}. Even though this is a very promising approach to 
a systematic description of odd-mass nuclei, the fact that several blocked states have to be considered 
at each deformation, as well as the explicit breaking of time-reversal symmetry, presents significant 
difficulties in realistic applications, especially for heavy nuclei. 

A wealth of new data on spectroscopic properties of odd-A nuclei in recent years has led to 
a renewed interest in particle-vibration coupling (PVC) approaches \cite{BM} that explicitly 
consider the polarization of a nucleus by the odd-particle. A number of PVC models of 
various levels of refinement and self-consistency 
have been developed \cite{litvinova2007,yoshida2009,colo2010,litvinova2011,mizuyama2012,tarpanov2014a,tarpanov2014b,niu2015} 
and applied to studies of structure phenomena. 
In this work we present an approach based on nuclear DFT and the 
particle-core coupling scheme \cite{BM,bohr1953}. It is an extension of a method 
introduced in Ref.~\cite{Nom08} for determining the Hamiltonian of the interacting boson model 
(IBM) \cite{IBM}, starting with an EDF-based SCMF calculation of 
deformation energy surfaces. By mapping a deformation constrained 
energy surface onto the equivalent Hamiltonian of the
IBM, that is, onto the energy expectation value in the
boson condensate state, the Hamiltonian parameters are determined. The 
resulting IBM Hamiltonian is used to calculate excitation spectra and transition 
rates. For an odd-mass nucleus this method is here applied to the even-even core, 
that is, the even-even core is described in terms of boson degrees of freedom, 
and only the fermion degrees of freedom of the odd unpaired particle(s) are 
treated explicitly. By extending the method of Ref.~\cite{Nom08} to systems 
with odd $N$ or/and $Z$, it becomes equivalent to the well known phenomenological 
interacting boson-fermion model (IBFM) \cite{IBFM,IBFM-Book}. The advantage of the 
present approach is that, except for the strength parameter(s) of the boson-fermion 
coupling, all parameters of the model Hamiltonian are determined by the choice of the 
(microscopic) energy density functional and pairing interaction. At the cost of having to 
adjust the boson-fermion coupling strength to data, we are able to include all states in 
a major shell in the fermion space, and extend the applicability of the approach to 
arbitrary heavy odd $N$ or/and $Z$ nuclei.

In Sec.~\ref{sec:model} we describe the method that will be used to determine the
boson-core and boson-fermion Hamiltonians, and compare the resulting parameters
with those of previous phenomenological calculation in
Ref.~\cite{scholten1982}. Sec.~\ref{sec:results} presents model results  
for spectroscopic properties, that is, excitation spectra and
electromagnetic transitions and moments. Sec.~\ref{sec:summary}
contains a short summary and concluding remarks. 


\section{Theoretical framework\label{sec:model}}

The model Hamiltonian for an odd-mass nucleus contains a term that corresponds to the 
even-even (boson) core $\hat H_{B}$, a single-particle Hamiltonian that describes the 
unpaired nucleon(s) $\hat H_F$, and a term that describes the interaction between 
bosons and fermions $\hat H_{BF}$:
\begin{eqnarray}
\label{eq:ibfm}
 \hat H = \hat H_B + \hat H_F + \hat H_{BF}. 
\end{eqnarray}
The assumption is that the Hamiltonian conserves separately the number of bosons 
$N_B$ and the number of fermions $N_F$. Here we will only consider the simplest case 
with $N_F =1$. For low-energy states, the dominant components in the boson space are the $s$
(spin $0^+$) and $d$ (spin $2^+$) bosons \cite{OAI}. The number of bosons equals the 
number of valence (spherical open-shell) proton and neutron pairs (particle or hole pairs). 
Since in the present study the model will be applied to axially-deformed rotational nuclei, 
for the boson Hamiltonian $\hat H_B$ we employ the standard form \cite{IBM}: 
\begin{eqnarray}
\label{eq:ibm}
 \hat H_B = \epsilon_d\hat n_d + \kappa\hat Q_B\cdot\hat Q_B +
  \kappa^{\prime}\hat L\cdot\hat L, 
\end{eqnarray}
with the $d$-boson number operator $\hat n_d=d^{\dagger}\cdot\tilde d$,
the quadrupole operator $\hat Q_B=s^{\dagger}\tilde d+d^{\dagger}\tilde s +
\chi[d^{\dagger}\times\tilde d]^{(2)}$, and the angular momentum
operator $\hat L=\sqrt{10}[d^{\dagger}\times\tilde d]^{(1)}$. 
$\epsilon_d$, $\kappa$, $\kappa^{\prime}$ and $\chi$ are parameters 
that will be determined by a DFT-based SCMF calculation. 
The fermion Hamiltonian for a single nucleon reads $\hat
H_F=\sum_{j}\epsilon_j[a^{\dagger}_j\times\tilde a_j]^{(0)}$, 
with $\epsilon_j$ the single-particle energy of the spherical orbital $j$. 
For the particle-core coupling $\hat H_{BF}$ we use the simplest form \cite{IBFM,IBFM-Book}: 
\begin{eqnarray}
\label{eq:bf}
 \hat H_{BF}=&&\sum_{jj^{\prime}}\Gamma_{jj^{\prime}}\hat
  Q_B\cdot[a^{\dagger}_j\times\tilde a_{j^{\prime}}]^{(2)} \nonumber \\
&&+\sum_{jj^{\prime}j^{\prime\prime}}\Lambda_{jj^{\prime}}^{j^{\prime\prime}}
:[[d^{\dagger}\times\tilde a_{j}]^{(j^{\prime\prime})}
\times
[a^{\dagger}_{j^{\prime}}\times\tilde d]^{(j^{\prime\prime})}]^{(0)}:
\nonumber \\
&&+\sum_j A_j[a^{\dagger}\times\tilde a_{j}]^{(0)}\hat n_d, 
\end{eqnarray}
where the first and second terms are referred to as the quadrupole and
exchange interaction, respectively. The former describes the quadrupole 
boson-fermion interaction, whereas the latter takes into account the fact that 
bosons are fermion pairs and its inclusion is essential for a detailed reproduction of low-energy 
spectra and electromagnetic transition probabilities. 
$\hat Q_B$ is the same boson quadrupole operator as in $\hat H_B$, and 
$:(\cdots):$ indicates normal ordering. 
The strength parameters $\Gamma_{jj^{\prime}}$ and
$\Lambda_{jj^{\prime}}^{j^{\prime\prime}}$ can be rewritten, by use of
the generalized seniority scheme, in the following $j$-dependent forms \cite{scholten1985}: 
\begin{eqnarray}
\label{eq:bf-strength}
&&\Gamma_{jj^{\prime}}=\Gamma_0\gamma_{jj^{\prime}} \\
&&\Lambda_{jj^{\prime}}^{j^{\prime\prime}}=-2\Lambda_0\sqrt{\frac{5}{2j^{\prime\prime}+1}}\beta
_{jj^{\prime}}\beta_{j^{\prime}j^{\prime\prime}}
\end{eqnarray}
where
$\gamma_{jj^{\prime}}=(u_ju_{j^{\prime}}-v_jv_{j^{\prime}})Q_{jj^{\prime}}$
and 
$\beta_{jj^{\prime}}=(u_jv_{j^{\prime}}+v_ju_{j^{\prime}})Q_{jj^{\prime}}$,
and the matrix element of the quadrupole operator in the single-particle
basis $Q_{jj^{\prime}}=\langle j||Y^{(2)}||j^{\prime}\rangle$. 
The factors $u_j$ and $v_j$ denote the occupation probabilities of the
orbit $j$, and satisfy $u_j^2+v_j^2=1$. 
The last term in Eq.~(\ref{eq:bf}) denotes the monopole term. $A_j$ can
be parametrized as $A_j=-\sqrt{2j+1}A_0$ with $A_0$ denoting the strength
parameter \cite{scholten1982}. 
The effect of this term is to  
compress or expand the single-nucleon energy levels \cite{IBFM}. 

\begin{figure}[htb!]
\begin{center}
\includegraphics[width=8.6cm]{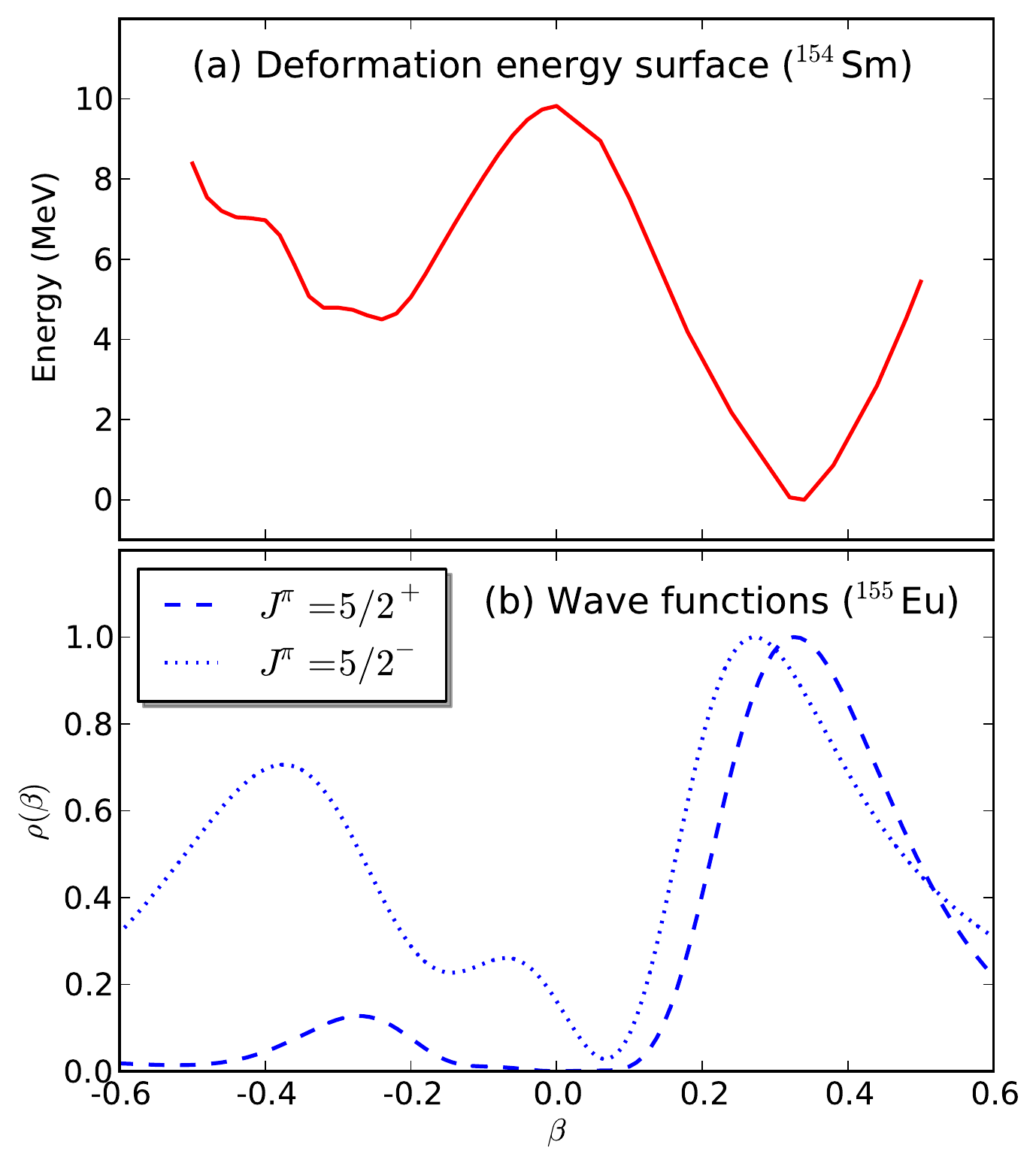} 
\caption{(Color online) (a) Projection of the RHB deformation energy surface along the 
$\beta$ axis for $^{155}$Eu, plotted with respect to the minimum. The
 calculation is based on the relativistic density functional 
DD-PC1 and  a separable pairing force of finite range. (b) 
Distribution $\rho(\beta)$ of the wave functions for the
 lowest-lying positive-parity ($J^{\pi}={5/2}^+$) and negative-parity 
 ($J^{\pi}={5/2}^-$) states of $^{155}$Eu (see the text for details). }
\label{fig:pes}
\end{center}
\end{figure}

As an illustrative application of the method, we consider the case of a single nucleon 
coupled to an axially-symmetric rotor: the low-energy spectra of the isotopes 
 $^{151,153,155}$Eu. These nuclei were extensively investigated 
in the IBFM calculation of Ref.~\cite{scholten1982} and, therefore, one can directly 
compare the present results with those obtained in a purely phenomenological approach. 
The corresponding even-even core nuclei $^{150,152,154}$Sm 
present excellent examples of axially-deformed rotors \cite{IBM}. 
The number of bosons equals the number of nucleon pairs outside the
doubly-magic core $^{132}$Sn, that is, 9, 10 and 11 for
$^{151,153,155}$Eu, respectively.

In the first step of the construction of the boson-fermion Hamiltonian $\hat H$ in
Eq.~(\ref{eq:ibfm}), the parameters for the even-even
core $\hat H_B$ Eq.~(\ref{eq:ibm}) are determined. 
To this aim we employ the procedure developed in Refs.~\cite{Nom08,nom11rot}: 
based on a specific choice for a nuclear EDF, the constrained SCMF
calculation determines the microscopic deformation energy surface as 
function of the polar deformation parameters $\beta$ and
$\gamma$ \cite{BM}.  This energy surface is mapped onto the 
corresponding expectation value of the boson Hamiltonian in the intrinsic (coherent) state
\cite{GK} of the interacting boson system, and this mapping completely determines 
the parameters of $\hat H_B$. 
Only the strength parameter $\kappa^{\prime}$ for the $\hat L\cdot\hat L$
term is determined separately so that the cranking moment of inertia in
the IBM intrinsic state becomes equal to the Inglis-Belyaev moment of
inertia ${\cal I}$ obtained from the self-consistent cranking
calculation at the mean-field minimum \cite{nom11rot}. 
Here ${\cal I}$ is increased by 30 \% taking into the fact that the
Inglis-Belyaev formula gives significantly smaller moment of inertia
than the empirical values. 

As an illustration, 
Fig.~\ref{fig:pes}(a) displays the projection of the energy map of 
$^{154}$Sm along the axis of $\beta$ deformation, obtained from the constrained self-consistent
relativistic Hartree-Bogoliubov (RHB) calculation based on the energy density functional DD-PC1
\cite{DDPC1} and a separable pairing force of finite range \cite{tian09}. 
One notices a pronounced prolate equilibrium minimum at $\beta_{\rm min} \approx 0.34$. 
The corresponding parameters $\epsilon_d$, $\kappa$, $\kappa^{\prime}$ and $\chi$ of $\hat H_B$, 
determined by the mapping of the microscopic energy surface, are summarized in Table
\ref{tab:parameter}.

\begin{table}[hb!]
\caption{\label{tab:parameter} Parameters of the boson Hamiltonian $\hat H_B$
 ($\epsilon_d$, $\kappa$, $\kappa^{\prime}$ and $\chi$), and single-particle 
 energies of the positive-parity orbits $2d_{5/2}$, $2d_{3/2}$ and
 $3s_{1/2}$, relative to the $1g_{7/2}$ orbit. All entries, except the
 dimensionless parameter $\chi$, are in MeV. }
\begin{center}
\begin{tabular*}{\columnwidth}{p{1.0cm}p{1.0cm}p{1.0cm}p{1.0cm}p{1.0cm}p{1.0cm}p{1.0cm}p{1.0cm}}
\hline\hline
\textrm{$N$} &
\textrm{$\epsilon_{d}$}&
\textrm{$\kappa$}&
\textrm{$\kappa^{\prime}$}&
\textrm{$\chi$}&
\textrm{$\epsilon_{d5/2}$}&
\textrm{$\epsilon_{d3/2}$}&
\textrm{$\epsilon_{s1/2}$}\\
\hline
88 & 0.46 & -0.079 & -0.017 & -0.55 & 3.14 & 5.04 & 5.74 \\
90 & 0.29 & -0.079 & -0.021 & -0.55 & 3.24 & 5.11 & 5.88 \\
92 & 0.13 & -0.079 & -0.022 & -0.55 & 3.32 & 5.14 & 5.98 \\
\hline\hline
\end{tabular*}
\end{center}
\end{table}

For the fermion valence space we include all the spherical single-particle orbits in the 
proton major shell $Z=50 - 82$: $1g_{7/2}$, $2d_{5/2}$, $2d_{3/2}$ and
$3s_{1/2}$ for positive-parity, and $1h_{11/2}$ 
for negative-parity, with single-particle (canonical) energies and occupation probabilities 
determined by the self-consistent RHB calculation constrained at zero deformation. 
It remains then to adjust the three strength parameters of the boson-fermion interaction 
Hamiltonian $\hat H_{BF}$. $\Gamma_0$, $\Lambda_0$ and $A_{0}$ are the only 
parameters that are fitted to data, separately for positive- and
negative-parity states. 
For each nucleus, the optimal values for the strength parameters are
chosen so as to reproduce the energies of the first excited state and of
the band-head of the second-lowest band. 
We include the monopole term only for $2d_{5/2}$ and $1h_{11/2}$ orbitals so
as to improve the description of the band-head of the second-lowest band.

The resulting total boson-fermion Hamiltonian $\hat H$ is diagonalized numerically in the spherical basis
$|j,L,\alpha,J\rangle$, where $\alpha$ is a generic notation for a set of quantum numbers
$n_d,\nu,n_{\Delta}$ that distinguish states with the same 
boson angular momentum $L$ \cite{IBM}, and $J$ is the total angular momentum of
the Bose-Fermi system ($|L-j|\leq J\leq L+j$).

To illustrate the method, we display in Fig.~\ref{fig:pes}(b) the distribution $\rho(\beta)$
of the wave functions for the lowest positive- and negative-parity states of $^{155}$Eu as 
functions of the axial deformation $\beta$. The function $\rho(\beta)$ is computed by taking 
the overlap between the eigenstate of
the IBFM Hamiltonian and the projected intrinsic state of the coupled
boson-fermion system expanded in terms of the basis
$|j,L,\alpha,J\rangle$ \cite{leviatan1988}. Starting from the spherical single-proton states, 
as a result of the interaction $\hat H_{BF}$ between the unpaired proton and the deformed 
boson core, the distributions of wave functions for both states $J^{\pi}={5/2}^+_1$
and ${5/2}^-_1$ display peaks close to the minimum of the energy surface of the 
even-even core $^{154}$Sm. The additional peaks at the corresponding negative values 
of $\beta$ ($\gamma = 60^\circ$) arise because the energy surface exhibits a parabolic 
dependence on $\gamma$ (cf. panel (a) of Fig.~\ref{fig:pes}).

\begin{figure}[htb!]
\begin{center}
\includegraphics[width=8.6cm]{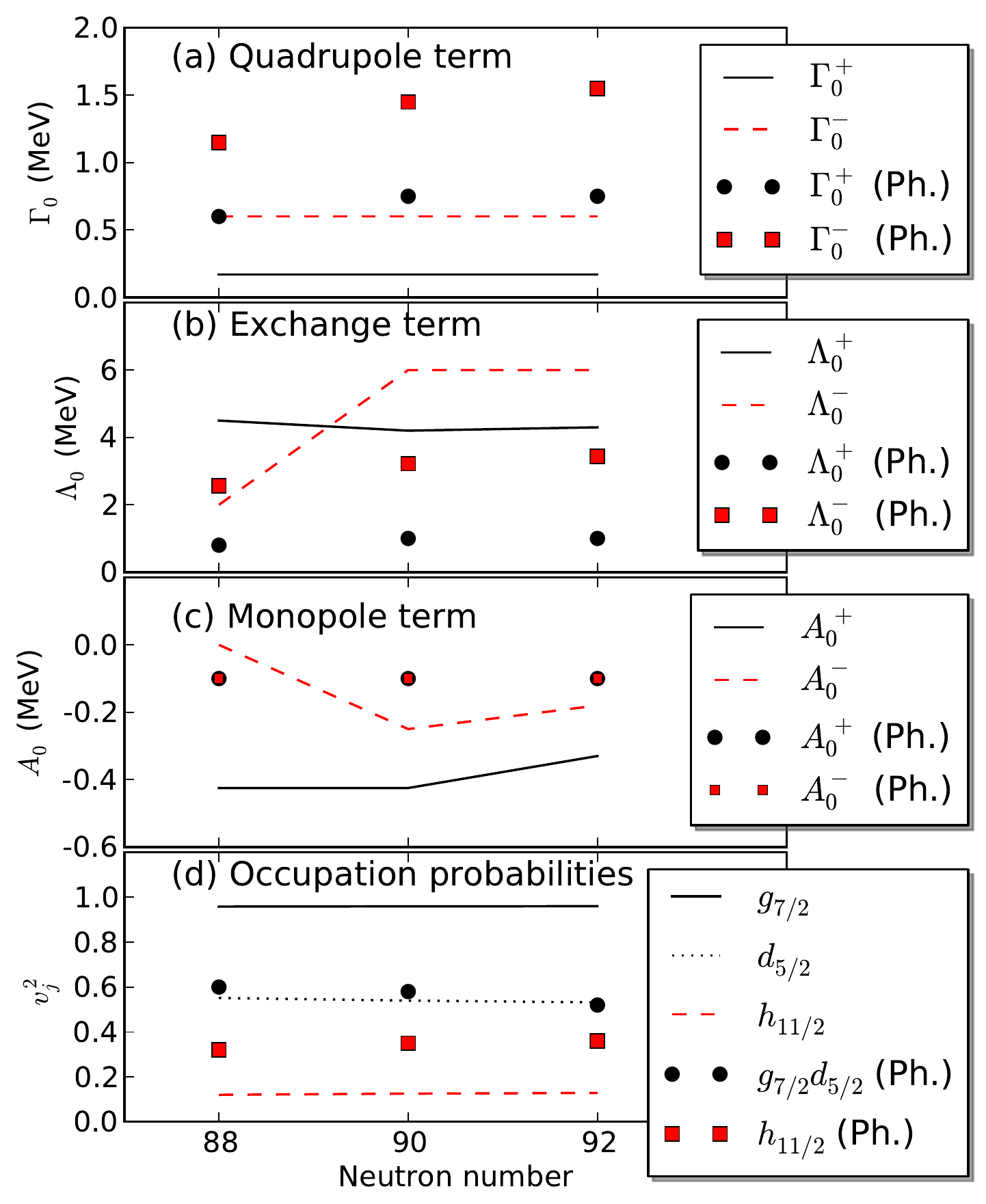} 
\caption{(Color online) Values of the parameters $\Gamma_0^{\pm}$ (a), 
 $\Lambda_0^{\pm}$ (b), $A_0^{\pm}$ (c) for positive- and
 negative-parity states, and the occupation probabilities $v^2$ of the spherical orbitals $1g_{7/2}$,
 $2d_{5/2}$ and $1h_{11/2}$ (d). 
 The corresponding values used in Ref.~\cite{scholten1982}, denoted by ``Ph.'', are shown for comparison. 
 In Ref.~\cite{scholten1982} identical occupation probabilities were used for the orbitals $1f_{7/2}$ and
 $2d_{5/2}$. }
\label{fig:para}
\end{center}
\end{figure}


In Fig.~\ref{fig:para} we plot the strengths parameters $\Gamma_0$ (a), 
$\Lambda_0$ (b), $A_{0}$ (c), and the RHB occupation probabilities $v^2_j$ of the spherical orbitals $1g_{7/2}$,
 $2d_{5/2}$ and $1h_{11/2}$ (d). The values used in the fully-phenomenological calculation
of Ref.~\cite{scholten1982} are also included for comparison.  
One notices that, for states of both parities, the values of 
$\Gamma_0$, $\Lambda_0$ and $A_{0}$ used in the present calculation are  
significantly different from the ones of Ref.~\cite{scholten1982}. 
The difference most likely results from the occupation probabilities
and energy spacing between the $1g_{7/2}$ and $2d_{5/2}$ single-particle
levels: from Fig.~\ref{fig:para}(d), $v_{g7/2}^2\approx 0.96$, $v_{d5/2}^2\approx 0.55$ and
$v^2_{h11/2}\approx 0.13$ in the present study, 
whereas $v_{g7/2}^2=v^2_{d5/2}\approx 0.5-0.6$ and $v^2_{h11/2}\approx
0.35$ in \cite{scholten1982}; 
From Tab.~\ref{tab:parameter}, we note that $|\epsilon_{g7/2}-\epsilon_{d5/2}|\approx 3$ MeV, 
in contrast to $< 0.5$ MeV in Ref.~\cite{scholten1982}. 
Nevertheless, in both studies $\Gamma_0$ and $\Lambda_0$ display a smooth variation 
with neutron number. Another difference is that in 
Ref.~\cite{scholten1982} the monopole boson-fermion interaction with a constant strength parameter $A_0=-0.1$
MeV was included in the Hamiltonian for all orbitals, whereas the
monopole term is used only for $2d_{5/2}$ and $1h_{11/2}$ orbitals and varies smoothly with
neutron number in the present calculation.

\section{Odd-A europium isotopes\label{sec:results}}

\begin{figure*}[htb!]
\begin{center}
\includegraphics[width=12.0cm]{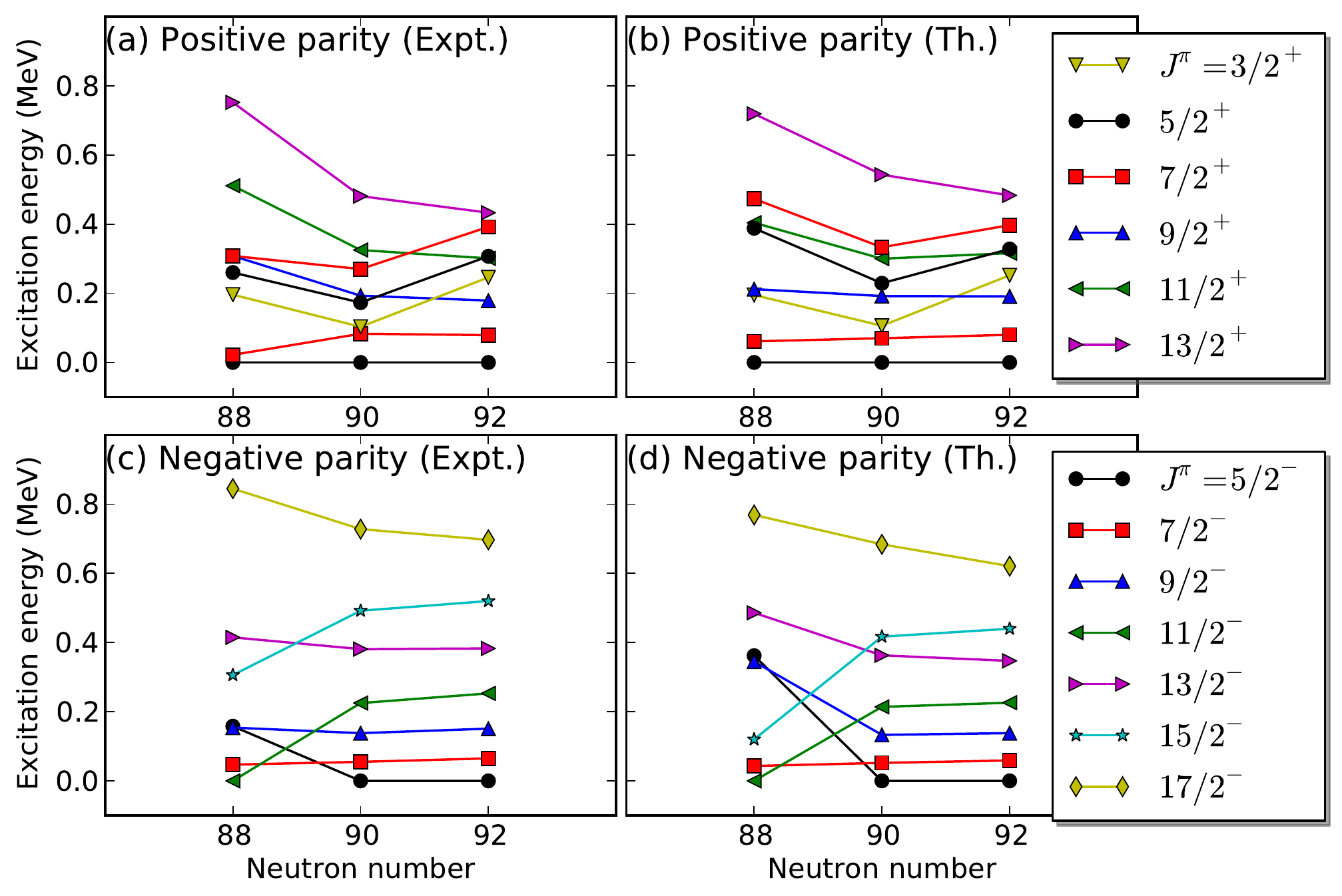} 
\caption{(Color online) The calculated low-energy positive- and negative-parity levels of $^{151,153,155}$Eu plotted  
in comparison with their experimental counterparts \cite{data}. }
\label{fig:level}
\end{center}
\end{figure*}

\begin{figure}[htb!]
\begin{center}
\includegraphics[width=8.6cm]{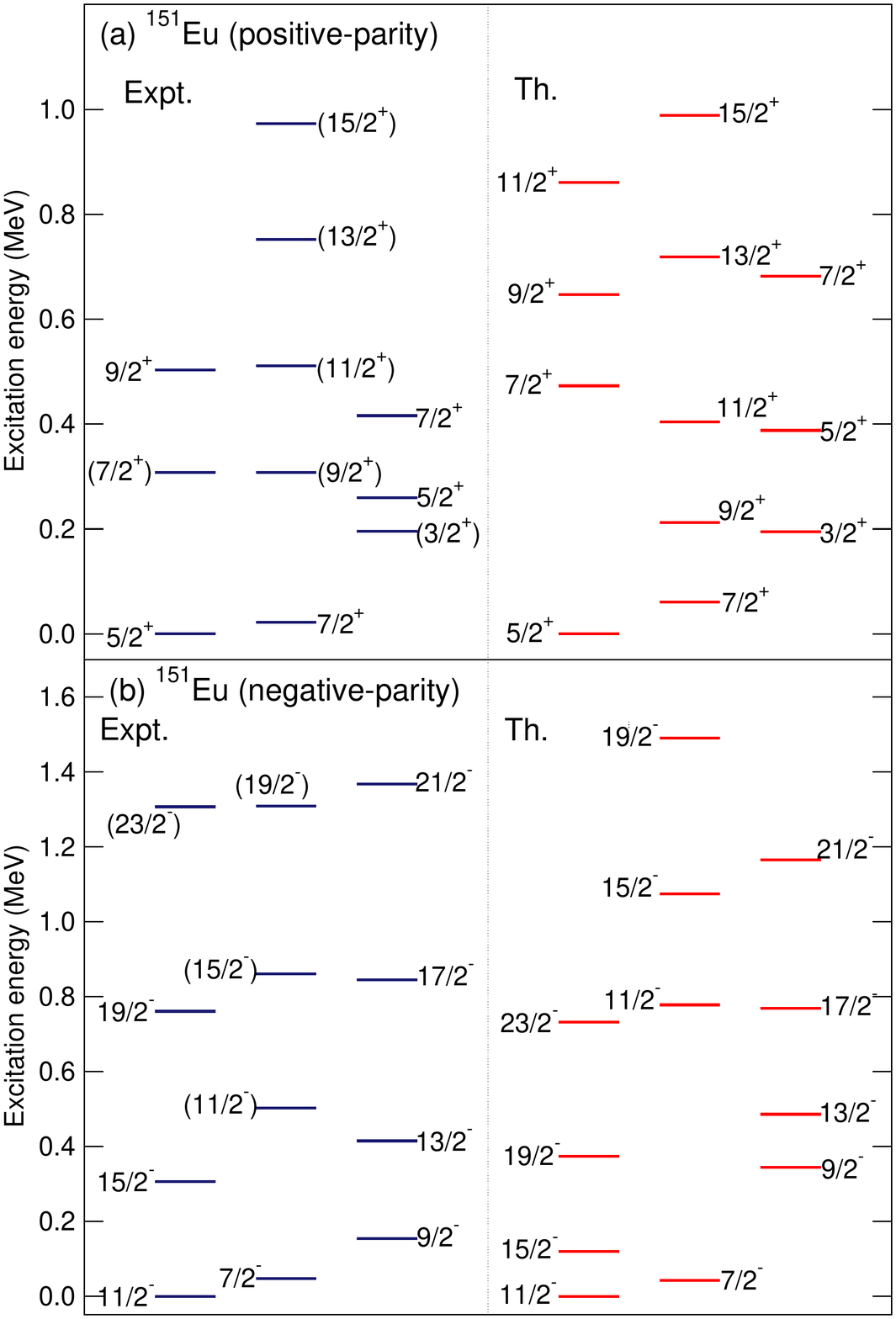} 
\caption{(Color online) Detailed comparison of the calculated low-energy levels of 
 $^{151}$Eu with available data \cite{data}. The excitation energies for
 the negative-parity states are shown relative to the lowest state. Note
 that the experimental states in parentheses denote tentative assignments.}
\label{fig:eu151}
\end{center}
\end{figure}

\begin{figure}[htb!]
\begin{center}
\includegraphics[width=8.6cm]{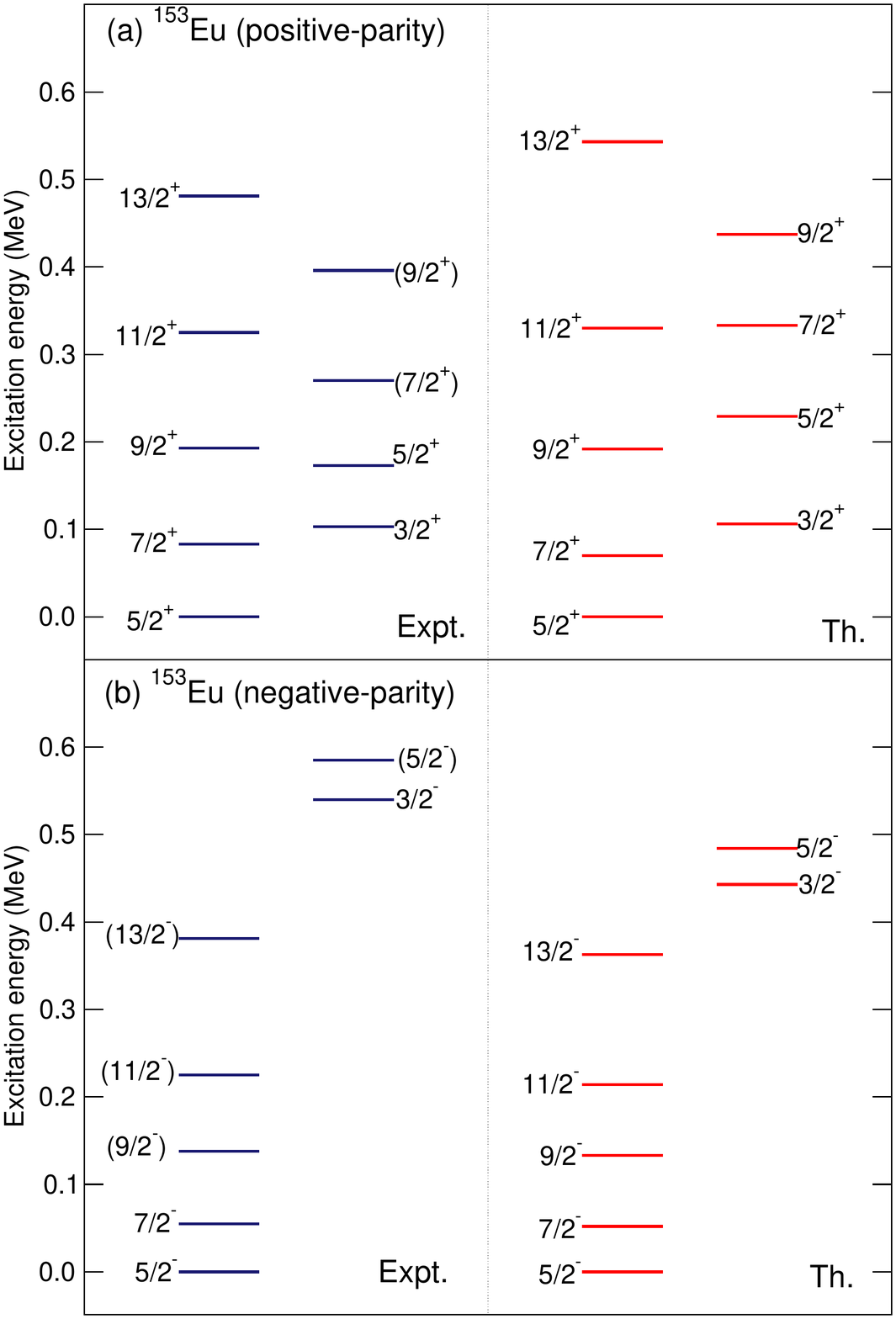} 
\caption{(Color online) Same as in the caption to Fig.~\ref{fig:eu151}, but
 for the isotope $^{153}$Eu. }
\label{fig:eu153}
\end{center}
\end{figure}

\begin{figure}[htb!]
\begin{center}
\includegraphics[width=8.6cm]{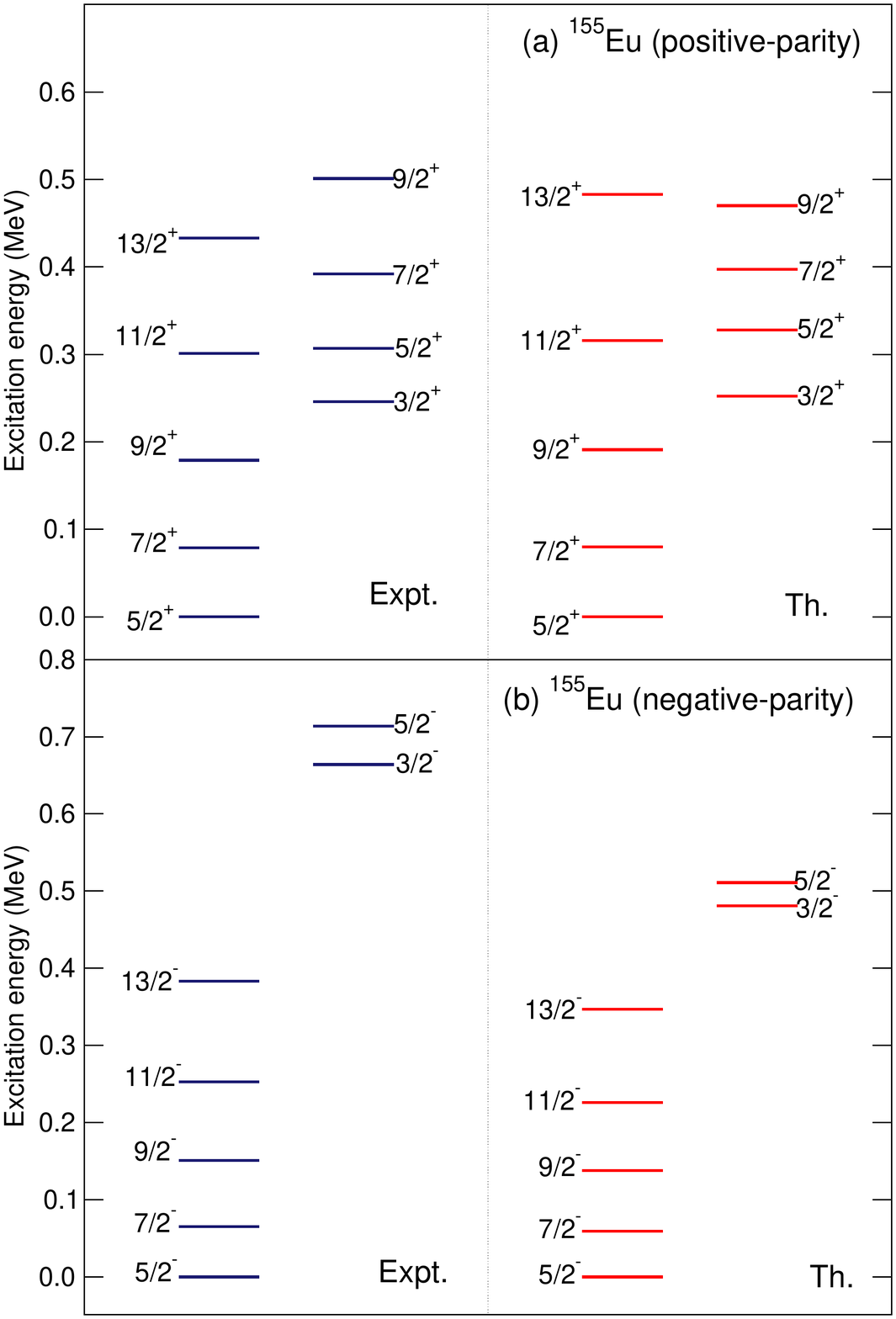} 
\caption{(Color online) Same as the caption to Fig.~\ref{fig:eu151}, but
 for $^{155}$Eu. }
\label{fig:eu155}
\end{center}
\end{figure}

\begin{table*}[b!]
\caption{\label{tab:trans} The calculated reduced E2 (in Weisskopf units)
 and M1 (in $\mu_N^2$) transition probabilities for  
 low-lying states, and spectroscopic quadrupole moments $Q(J^{\pi})$
 (in $b$) and 
 magnetic moments $\mu(J^{\pi})$ (in $\mu_N$) for $^{151,153,155}$Eu, 
 compared to available experimental values \cite{data,hess1970,guenther1976,kroger1975,stone2005}.}
\begin{center}
\begin{tabular}{p{5.0cm}cccccc}
\hline\hline
\multirow{2}{*}{} & \multicolumn{2}{c}{$^{151}$Eu} & \multicolumn{2}{c}{$^{153}$Eu} & \multicolumn{2}{c}{$^{155}$Eu} \\
\cline{2-3} 
\cline{4-5}
\cline{6-7}
          & Th.         & Expt.    & Th.         & Expt.     & Th.     & Expt. \\
\hline
$B({\textnormal{E2}};{3/2}^+_1\rightarrow{5/2}^+_1)$ 
& 18 & 22(12) & 40 & 1.4(5) & 4.3 & 0.47(9)\\
$B({\textnormal{E2}};{3/2}^+_1\rightarrow{7/2}^+_1)$ 
& 4.0 & 2.4(4) & 22 & 1.9(4) & 3.6 & 0.73(6)\\
$B({\textnormal{E2}};{5/2}^+_2\rightarrow{3/2}^+_1)$ 
& 86 & - & 210 & 154(82) & 281 & - \\
$B({\textnormal{E2}};{5/2}^+_2\rightarrow{5/2}^+_1)$ 
& 10 & - & 34 & 0.9(4) & 3.9 & - \\
$B({\textnormal{E2}};{5/2}^+_2\rightarrow{7/2}^+_1)$ 
& 0.8 & - & 0.8 & 2.8(1.7) & 0.012 & - \\
$B({\textnormal{E2}};{7/2}^+_1\rightarrow{5/2}^+_1)$ 
& 57 & 8.1(9) & 190 & 300(21) & 267 & - \\
$B({\textnormal{E2}};{7/2}^+_2\rightarrow{5/2}^+_1)$ 
& 26 & - & 7.2 & 0.53(7) & 9.8 & - \\
$B({\textnormal{E2}};{7/2}^+_2\rightarrow{7/2}^+_1)$ 
& 34 & $<$80 & 47 & 4(4) & 24 & - \\
$B({\textnormal{E2}};{9/2}^+_1\rightarrow{5/2}^+_1)$ 
& 10 & - & 51 & 97(8) & 68 & - \\
$B({\textnormal{E2}};{9/2}^+_1\rightarrow{7/2}^+_1)$ 
& 87 & - & 198 & 179(21) & 246 & - \\ 
$B({\textnormal{E2}};{3/2}^-_1\rightarrow{5/2}^-_1)$ 
& 11 & - & 20 & - & 0.5 & - \\
$B({\textnormal{E2}};{7/2}^-_1\rightarrow{5/2}^-_1)$ 
& 38 & - & 241 & - & 266 & - \\
$B({\textnormal{E2}};{9/2}^-_1\rightarrow{7/2}^-_1)$ 
& 6.7 & $>$70 & 197 & - & 222 & - \\
$B({\textnormal{M1}};{3/2}^+_1\rightarrow{5/2}^+_1)$ 
& 0.0098 & 0.0078(16)& 0.016 & 0.0034(1) & 0.000072 & 0.00098(9)\\
$B({\textnormal{M1}};{5/2}^+_2\rightarrow{3/2}^+_1)$ 
& 0.167& - & 0.11 & 0.22(2) & 0.120 & - \\
$B({\textnormal{M1}};{5/2}^+_2\rightarrow{5/2}^+_1)$ 
& 0.00024 & - & 0.00055 & 0.00016(4) & 0.00030 & - \\
$B({\textnormal{M1}};{5/2}^+_2\rightarrow{7/2}^+_1)$ 
& 0.0077  & - & 0.012 & 0.0030(3) & 0.0011 & - \\
$B({\textnormal{M1}};{7/2}^+_1\rightarrow{5/2}^+_1)$ 
& 0.0060 & 0.0083(4) & 0.020 & 0.011(1) & 0.021 & - \\
$B({\textnormal{M1}};{7/2}^+_2\rightarrow{5/2}^+_1)$ 
& 0.089 & 0.20(5) & 0.048 & 4.6$\times 10^{-5}$(4) & 0.046 & - \\
$B({\textnormal{M1}};{7/2}^+_2\rightarrow{7/2}^+_1)$ 
& 0.020 & 0.015(5) & 0.018 & 0.00106(6) & 0.011 & - \\
$B({\textnormal{M1}};{9/2}^+_1\rightarrow{7/2}^+_1)$ 
& 0.016 & - & 0.034 & 0.016(1) & 0.033 & - \\
$B({\textnormal{M1}};{3/2}^-_1\rightarrow{5/2}^-_1)$ 
& 1.73 & - & 2.49 & - & 2.75 & - \\
$B({\textnormal{M1}};{7/2}^-_1\rightarrow{5/2}^-_1)$ 
& 0.26 & - & 0.014 & - & 0.14 & - \\
$B({\textnormal{M1}};{9/2}^-_1\rightarrow{7/2}^-_1)$ 
& 0.55 & - & 0.039 & - & 0.20 & - \\
$Q({3/2}^+_1)$ & +0.70 & - & +1.14 & 1.254(13) & +1.27 & - \\
$Q({5/2}^+_1)$ & +1.16 & +0.903(10) & +1.79 & +2.28(9) & +2.26 &
 +2.49(2) \\
$Q({7/2}^+_1)$ & +0.79 & 1.28(2) & +0.63 & +0.44(2) & +0.63 & - \\
$\mu({3/2}^+_1)$ & +1.22 & - & +1.25 & +2.048 & +1.22 & - \\
$\mu({5/2}^+_1)$ & +1.53 & +3.4717(6) & +1.54 & +1.53  & +1.52 &
 +1.52 \\
$\mu({7/2}^+_1)$ & +2.02 & +2.591(2) & +1.93 & +1.81(6) & +1.86 & - \\
$\mu({5/2}^-_1)$ & +5.52 & - & +3.05 & +3.22(23)  & +2.96 & +9.6(10) \\
\hline\hline
\end{tabular}
\end{center}
\end{table*}

Figure \ref{fig:level} compares the calculated 
low-energy positive- and negative-parity levels of $^{151,153,155}$Eu 
to available data \cite{data}. 
For the positive-parity states (Fig.~\ref{fig:level} (a,b)), the ${5/2}^+_1$ 
state is the ground state in all three nuclei.  
For $^{153,155}$Eu the levels above the ground state, that is, ${7/2}^+_1$,
${9/2}^+_1$, ${11/2}^+_1$ and ${13/2}^+_1$, form a rotational
band with excitation energies proportional to $J(J+1)$ (cf. also
Figs.~\ref{fig:eu153} and \ref{fig:eu155}). 
$^{151}$Eu differs in structure from 
$^{153,155}$Eu by the fact that its ${7/2}^+_1$ state is low in energy
and close to the ${5/2}^+_1$ ground state. 
Indeed, its boson core $^{150}$Sm 
is rather close to the transitional region between rotational and vibrational
nuclei, whereas $^{153,155}$Eu are prolate deformed rotors. 
Considering that only three free parameters are adjusted to data, the 
calculation quantitatively reproduces the experimental systematics, except
perhaps for the excitation energy of the ${7/2}^+_2$ level in $^{151}$Eu. 
The ${3/2}^+_1$ state is supposed to be the bandhead of the excited band,
followed by the levels ${5/2}^+_2$ and ${7/2}^+_2$. 
For the negative-parity states in Fig.~\ref{fig:level} (c,d), the model results 
agree with the empirical rotational-like level structure in $^{153,155}$Eu. 
A significant structural change is obtained in $^{151}$Eu, in which the
${11/2}^-$ level becomes the ground state.

We emphasize the fact that the model can describe not only systematic trends
of low-lying levels, but also details of excitation spectra and decay patterns
in individual nuclei. Figures~\ref{fig:eu151}, \ref{fig:eu153} and \ref{fig:eu155} 
display the comparison between
theoretical and experimental low-energy levels for the positive and
negative-parity states of $^{151,153,155}$Eu, respectively. 
The levels are grouped into bands according to the dominant decay
pattern. One notes that, overall, the theoretical results are in good
agreement with experiment, particularly for the more deformed $^{153,155}$Eu. 
The present results reproduce data on the same level of accuracy as the 
fully phenomenological approach of Ref.~\cite{scholten1982}. 

For $^{151}$Eu it was suggested in Ref.~\cite{scholten1982} that the positive-parity ${5/2}^+_1$ and
${7/2}^+_1$ states correspond to the $2d_{5/2}$ and $1g_{7/2}$ single-particle
states, and become the band-heads of the bands $({5/2}^+_1,{7/2}^+_2,{9/2}^+_2,\ldots)$ and 
$({7/2}^+_1,{9/2}^+_1,{11/2}^+_1,\ldots)$, respectively. 
In the present study, in contrast, the levels that belong to both
bands, built on ${5/2}^+_1$ and
${7/2}^+_1$ states, are predominantly based on the $1g_{7/2}$
configuration. 
For the negative-parity states of $^{151}$Eu in Fig.~\ref{fig:eu151}(b)
our calculation predicts that the three bands based on ${11/2}^-_1$, ${7/2}^-_1$ and ${9/2}^-_1$ 
follow the $\Delta J=2$ systematics of states decoupled from the deformation of the core. 
Figures~\ref{fig:eu153} and \ref{fig:eu155} show that the band structures of $^{153}$Eu and
$^{155}$Eu are very similar. 
In both nuclei the two positive-parity bands built on the states
$J^{\pi}={5/2}^+$ and ${3/2}^+$ are assigned to the 
$K^{\pi}={5/2}^+$ and $K^{\pi}={3/2}^+$ rotational bands, respectively. 
The level energies of these $J(J+1)$ rotational bands exhibit the 
strong-coupling $\Delta J=1$ systematics. 
The positive-parity bands based on ${5/2}^+_1$ and ${3/2}^+_1$ predominantly correspond 
to the $1g_{7/2}$ and $2d_{5/2}$ proton configurations, respectively, with significant mixing 
of the two configurations. The similarity between band structures in $^{153}$Eu and
$^{155}$Eu is also evident for the negative-parity bands based on the $1h_{11/2}$ spherical orbital.

The relevant decay modes are the electric quadrupole (E2) and magnetic
dipole (M1) transitions. The corresponding  operator $T^{(E2)}=e_B\hat Q_B + e_F\hat Q_F$, with $\hat
Q_F=-\sum_{jj^{\prime}}\gamma_{jj^{\prime}}[a^{\dagger}_j\times\tilde
a_{j^{\prime}}]^{(2)}/\sqrt{5}$, and $e_B$ and $e_{F}$ denote the effective
charges. $e^B$ is adjusted to reproduce the experimental $B(E2;
2^+_1\rightarrow 0^+_1)$ value for the boson core nucleus, and the constant value
$e_F=1.0$ $e$b is used for the unpaired proton. 
The magnetic dipole operator $T^{(M1)}=\sqrt{3/4\pi}(g^B\hat L + \sum_{j}g^F_j\hat
j)$, where $\hat j$ is the fermion angular momentum operator, 
$g^B$ and $g^F_j$ are the boson and fermion $g$-factors,
respectively. $g^B$ is adjusted to the experimental magnetic moment of
the $2^+_1$ state of the boson core, $g^B=\mu(2^+_1)/2$, and the Schmidt
values are used for $g^F_j$, with the spin $g$-factor quenched by 30 \%. 
Data are available for E2 and M1 transitions 
between low-lying states of positive-parity bands. 

Table~\ref{tab:trans} collects the results for the E2 and M1 transition strengths, 
spectroscopic quadrupole moments and magnetic moments. 
In general, with only a few exceptions, the present study reproduces available data
\cite{data,hess1970,guenther1976}, and is consistent with the results
obtained in Ref.~\cite{scholten1982}. 
In this calculation rather strong in-band E2 and M1 transitions 
for the bands built on the ${5/2}^+_1$ and ${3/2}^+_1$ states are
predicted for $^{153,155}$Eu. Because of the pronounced mixing 
between the $2d_{5/2}$ and $1g_{7/2}$ configurations, 
the calculated inter-band transitions are rather large in all 
considered isotopes, and overestimate the data such as, for instance, the
${7/2}^+_1\rightarrow{5/2}^+_1$ and ${3/2}^+_1\rightarrow{5/2}^+_1$ E2
transitions in $^{151}$Eu and $^{153}$Eu, respectively. Note that,
except for $^{153}$Eu, the data on $B$(E2) and
$B$(M1) values are rather scarce. The calculated spectroscopic quadrupole and magnetic moments for 
low-lying states are in good agreement with the available experimental values \cite{data,stone2005}.

\section{Summary\label{sec:summary}}

In conclusion, we have introduced an advanced method for calculating 
spectroscopic properties of medium-mass and heavy nuclei with odd $N$ or/and $Z$.
The IBFM Hamiltonian used to describe the coupled system of the 
unpaired particle(s) plus boson-core, is based on the microscopic 
framework of nuclear energy density functional theory. 
The deformation energy surface of the even-even core, as well as the 
spherical single-particle energies and occupation probabilities of the odd
particle(s), are obtained in a SCMF calculation determined by the choice of the 
energy density functional and pairing interaction. Only the strength 
parameter(s) of the boson-fermion interaction Hamiltonian are 
specifically adjusted to data for each nucleus. As an illustrative 
example, the low-energy excitation spectra and transition rates 
of $^{151-155}$Eu have been analyzed, and a very good agreement with 
data has been obtained. The microscopic approach in which the 
even-even core is described in terms of boson degrees of freedom, and 
only the fermion degrees of freedom of the unpaired particle(s) are 
treated explicitly, enables an accurate, computationally feasible, and  
systematic description of a wealth of new data on isotopes with an 
odd number of protons or/and neutrons.

\begin{acknowledgements}
 
We thank N. Yoshida for providing us with the IBFM code PBOS. 
This work has been supported in part by 
the Croatian Science Foundation -- project ``Structure and Dynamics
of Exotic Femtosystems" (IP-2014-09-9159) and by the QuantiXLie Centre of Excellence.

\end{acknowledgements}

\bibliography{refs}

\end{document}